\documentstyle[preprint,aps]{revtex}
\begin{document}
\title{Relativity and the Minimum Slope of the Isgur-Wise Function}
\author{B. D. Keister}
\address{
  Department of Physics,
  Carnegie Mellon University,
  Pittsburgh, PA 15213}
\date{\today}
\maketitle
\begin{abstract}
  Sum rules based upon heavy quark effective theory indicate that the
  Isgur-Wise function $\xi(w)$ has a minimum slope $\rho_{\rm min}^2$
  as $w\to 1$, where $\rho_{\rm min}^2=0$ for light degrees of freedom
  with zero spin and $\rho_{\rm min}^2={1\over4}$ for light spin
  ${1\over2}$.  Quark-model studies reveal sources for a minimum slope
  from a variety of relativistic effects.  In this paper the origins
  of the minimum slope in the sum-rule and quark-model approaches are
  compared by considering hadrons with arbitrary light spin.  In both
  approaches the minimum slope increases with the light spin $j_l$,
  but there appears to be no detailed correspondence between the
  quark-model and sum-rule approaches.
\end{abstract}
\newpage
\narrowtext

The physics of hadrons containing a single heavy quark displays
additional symmetries beyond that of QCD~\cite{IW}.  Matrix elements
of operators which act only on the heavy quark may depend upon the
hadronic structure, but only trivially on the heavy quark mass in the
limit that it is infinitely heavy.  Transitions involving the heavy
quark quantum numbers can be expressed in terms of universal functions
$\xi(w)$, where $w=(v'\cdot v)$ and $v$ and $v'$ are the initial and
final hadron four-velocities.

On very general grounds, in heavy quark effective theory (HQET) a sum
rule can be derived~\cite{Bjorken,IWsum} in which the Isgur-Wise
function has the form $\xi(w) = 1 - \rho^2(w-1) + \cdots$ as $w\to 1$,
and that the slope $\rho^2$ has a minimum value of ${1\over4}$ for
heavy quarks coupled to light degrees of freedom with spin
${1\over2}$.  Isgur and Wise~\cite{IWsum} noted that this factor
differs from the value zero obtained in nonrelativistic calculations,
and also when the light spin $j_l=0$, and they further speculated that
it comes from Zitterbewegungen of the fermionic light degrees of
freedom.  The problem of a heavy quark coupled to light degrees of
freedom with arbitrary spin was later addressed by Falk~\cite{Falk},
who showed that for all light spins $j_l>0$, both integral and
non-integral, there is a non-vanishing minimum slope, thereby ruling
out Zitterbewegungen as the relevant phenomenon.

The minimum slope of $\xi(w)$ has been investigated in the context of
a quark model by Close and Wambach~\cite{Close}, who find $\rho=1.19$
for their choice of model parameters.  Le~Yaouanc {\it et
  al.}~\cite{LeYaouanc1,LeYaouanc2,LeYaouanc3} have examined
systematically the relativistic contributions to the minimum slope
using the framework of direct
interactions~\cite{BakamjianThomas,KeisterPolyzou}.  They find three
distinct sources: kinematics involving the bound wave function and the
Jacobian for variable transformations, a normalization effect in the
heavy quark current, and Wigner rotations of the light spectator.
These authors used instant-form dynamics, and concentrated on systems
where the light degrees of freedom have spin~${1\over2}$.

In this paper we compare the origins of the minimum slope in the HQET
sum-rule and quark-model approaches by considering systems where the
light degrees of freedom have arbitrary spin.  The only contribution
to the minimum slope which depends upon the light spin $j_l$ is the
effect of Wigner rotations; its contribution to the slope grows with
increasing $j_l$, a result also obtained in an analysis by Falk, who
uses a Minkowski tensor representation of the light degrees of
freedom, though the functional dependence upon $j_l$ is different.
However, it will also be seen that there are difficulties in making a
detailed correspondence between the physics leading to the HQET
sum-rule results and those of a model.

Our approach is similar to that of Le~Yaouanc {\it et al.}, in that we
use a Bakamjian-Thomas direct interaction~\cite{BakamjianThomas}, but
we employ Dirac's point-form dynamics rather than the instant form.
The advantage of this approach is that exposes the effects of
non-infinite heavy masses in a transparent manner.  In the heavy-quark
limit, all forms of dynamics should yield the same results.  The
details of this approach are described elsewhere~\cite{BKheavy}.

The plan of the paper is as follows: after a brief overview of the
dynamical approach, a heuristic derivation is given of the
contribution of Wigner rotations to the minimum slope for
spin-${1\over2}$ light degrees of freedom; then the exact results for
the minimum slope of Le~Yaouanc {\it et al.}\ are extended to
arbitrary spin and compared to those of the Minkowski tensor
approach. 

For calculational purposes, we employ non-covariant free-particle
state vectors $|m j; {\bf v} \mu\rangle $ for a particle of mass
$m$ and spin $j$, which are normalized as follows:
\begin{equation}
  \langle m j; {\bf v}' \mu'| m j; {\bf v} \mu\rangle  
  = \delta_{\mu'\mu}\delta({\bf v}' - {\bf v}).
  \label{AAA}
\end{equation}
We now consider matrix elements of the form $\langle M'j';
{\bf V}' \mu'| I^\alpha(0) | M j; {\bf V} \mu\rangle $,
where $I^\alpha(x)$ acts on the quantum numbers of the heavy quark.

In general, the heavy quark is bound to a light system with spin $j_l$
and mass $m_l$.  The matrix element is expressed in terms of
single-particle states as follows:
\begin{eqnarray}
  \langle M'j'; {\bf V}' \mu'| I^\alpha(0) | M j; {\bf V} \mu\rangle 
  &=& \int d{\bf v}_h' \int d{\bf v}_h \int d{\bf v}_l 
  \langle M'j'; {\bf V}' \mu' | m_h' m_l {\textstyle{1\over2}} j_l;
  {\bf v}_h' {\bf v}_l \mu_h' \mu_l \rangle  \\
  &&\quad\times
  \langle  m_h' {\textstyle{1\over2}};
  {\bf v}_h' \mu_h' | I^\alpha(0) |
  m_h {\textstyle{1\over2}}; {\bf v}_h 
  \mu_h \rangle
  \langle  m_h m_l {\textstyle{1\over2}} j_l;
  {\bf v}_h {\bf v}_l \mu_h \mu_l | M j; {\bf V} \mu \rangle \nonumber
  \label{AAH}
\end{eqnarray}
Here and henceforth repeated indices are summed.  The non-interacting
two-body mass of the system is
\begin{equation}
  M_0^2 \equiv P_0^\alpha P_{0\alpha}; 
  \qquad P_0^\alpha \equiv m_h v_h^\alpha + m_l v_l^\alpha.
  \label{AAD}
\end{equation}
The zero subscript denotes a non-interacting system.  The total
velocity is 
\begin{equation}
  {\bf V} = {\bf P} / M_0.
  \label{AADA}
\end{equation}
The relative momentum ${\bf k}$ can be expressed in terms of ${\bf
p}_l = m_l{\bf v}_l$ and a rotationless boost $L_c({\bf V})$:
\begin{eqnarray}
  {\bf k} &&= L_c^{-1}({\bf V}) {\bf p}_l \nonumber  \\
  &&\equiv {\bf p}_l - {\bf V}\left[ \omega_{m_l}({\bf p}_l)
    - { ({\bf V}\cdot{\bf p}_l)\over 1 + V^0}\right],
\label{AAF}
\end{eqnarray}
where 
\begin{equation}
  \omega_m({\bf p}) \equiv \sqrt{m^2 + {\bf p}^2}.
  \label{AAFA}
\end{equation}
In terms of ${\bf k}$, the non-interacting mass is
\begin{equation}
  M_0 = \omega_{m_h}({\bf k}) + \omega_{m_l}({\bf k}).
  \label{AAFAA}
\end{equation}

The most general bound state wave function is expressed as follows:
\begin{eqnarray}
  \langle  m_h m_l {\textstyle{1\over2}} j_l;
  {\bf v}_h {\bf v}_l \mu_h \mu_l | M j; {\bf V} \mu \rangle 
  &=& \left| {\partial({\bf V}{\bf k})
      \over \partial({\bf v}_h{\bf v}_l)} \right|^{1\over2}
  \delta({\bf V} - {\bf V}({\bf v}_h, {\bf v}_l))
  \nonumber \\
  &&\quad\times
  D^{{1\over2}{\dagger}}_{\mu_h{\bar\mu}_h}[R_c(L_c({\bf V}), 
  {\bf k}/m_h)]
  D^{j_l{\dagger}}_{\mu_l{\bar\mu}_l}[R_c(L_c({\bf V}),
  -{\bf k}/m_l)]
  \nonumber \\
  &&\quad\times
  \langle {\textstyle{1\over2}} {\bar\mu}_h 
  j_l {\bar\mu}_l | s \mu_s \rangle 
  \langle l \mu_l s \mu_s | j \mu \rangle 
  Y^L_{\mu_L}({\hat{\bf k}}) \phi_{L}(k),
  \label{AAI}
\end{eqnarray}
where the Jacobian is
\begin{equation}
  \left| {\partial({\bf v}_h{\bf v}_l) 
      \over\partial({\bf V}{\bf k})} \right|
  = \left({M_0^3 \over m_h^3 m_l^3}\right)
  {\omega_1({\bf v}_h)\omega_1({\bf v}_l)
    \over \omega_1({\bf u}_h)\omega_1({\bf u}_l)\omega_1({\bf V})};
  \qquad {\bf u}_{h,l} \equiv {\bf k} / m_{h,l},
  \label{AAE}
\end{equation}
and
\begin{equation}
  {\bf V}({\bf v}_h, {\bf v}_l)
  = {1\over M_0} (m_h{\bf v}_h + m_l {\bf v}_l)
  \label{AAJ}
\end{equation}
and $\phi_{L}(k)$ is a radial momentum wave function for an
interacting state of mass $M$.

At this point we convert from $LS$ to $jj$ coupling and assume that
the heavy quark is coupled to a ``light spin'' $j_l$ which contains
all angular momenta except for the spin of the heavy quark
itself~\cite{ISGW2}.  In that case we set $L=0$ and Eq.~\ref{AAE}
simplifies to 
\begin{eqnarray}
  \label{AAEA}
  \langle  m_h m_l {\textstyle{1\over2}} j_l
  {\bf v}_h {\bf v}_l \mu_h \mu_l | M j; {\bf V} \mu \rangle 
  &=& \left| {\partial({\bf V}{\bf k})
      \over \partial({\bf v}_h{\bf v}_l)} \right|^{1\over2}
  \delta({\bf V} - {\bf V}({\bf v}_h, {\bf v}_l))
  \nonumber \\
  &&\quad\times
  D^{{1\over2}{\dagger}}_{\mu_h{\bar\mu}_h}[R_c(L_c({\bf V}), 
  {\bf k}/m_h)]
  D^{j_l{\dagger}}_{\mu_l{\bar\mu}_l}[R_c(L_c({\bf V}),
  -{\bf k}/m_l)]
  \nonumber \\
  &&\quad\times
  \langle {\textstyle{1\over2}} {\bar\mu}_h 
  j_l {\bar\mu}_l | s \mu_s \rangle 
  (4\pi)^{-{1\over2}} \phi_{0}(k),
\end{eqnarray}

The Wigner rotation has the following
form:
\begin{equation}
  \label{AAJA}
  R_c(L_c({\bf v}), {\bf a}) = L_c^{-1}({\bf a}') L_c({\bf v})
  L_c({\bf a});\qquad a'{}^\mu = L_c^\mu{}_\nu({\bf v}) a^\nu.
\end{equation}

After some variable changes, we get
\begin{eqnarray}
\langle M'j'; {\bf V}' \mu'| I^\alpha(0) | M j; {\bf V} \mu\rangle  
&=&(4\pi)^{-1}\int d{\bf k} 
\left|{\partial {\bf V}' \over \partial {\bf v}'_h}
\right|_{{\bf v}_l}^{-1}
\left| {\partial({\bf v}_h{\bf v}_l) 
\over\partial({\bf V}{\bf k})} \right|^{1\over2}
\left| {\partial({\bf V}'{\bf k}')
\over\partial({\bf v}_h'{\bf v}_l') } \right|^{1\over2}\nonumber \\
&&\quad\times
\langle {\textstyle{1\over2}} {\bar\mu}_h' j_l
{\bar\mu}_l' | s' \mu_s' \rangle 
\phi_{0}^*(k') \nonumber  \\
&&\quad\times
D^{1\over2}_{{\bar\mu}_h'\mu_h'}[R_c(L_c({\bf V}'), {\bf k}'/m_h')]
D^{j_l}_{{\bar\mu}_l'\mu_l}[R_c(L_c({\bf V}'),-{\bf k}'/m_l)]
\nonumber \\
&&\qquad\qquad\times
\langle  m_h' {\textstyle{1\over2}}; {\bf v}_h' \mu_h' | I^\alpha(0)
| m_h {\textstyle{1\over2}}; {\bf v}_h \mu_h \rangle \nonumber \\
&&\quad\times
D^{{1\over2}{\dagger}}_{\mu_h{\bar\mu}_h}[R_c(L_c({\bf V}), {\bf k}/m_h)]
D^{j_l{\dagger}}_{\mu_l{\bar\mu}_l}[R_c(L_c({\bf V}),-{\bf k}/m_l)]
\nonumber \\
&&\quad\times
\langle {\textstyle{1\over2}} {\bar\mu}_h j_l
{\bar\mu}_l | s \mu_s \rangle 
\phi_{0}(k),
\label{AAS}
\end{eqnarray}
where
\begin{equation}
\left|{\partial {\bf V}' \over \partial {\bf v}'_h}\right|_{{\bf v}_l}
= \left({m_h' \over M_0'}\right)^3
\left[1 - {1\over v_h'{}^0} {\bf V}'\cdot 
(V'{}^0{\bf v}_h' - v_h'{}^0{\bf V}')\right].
\label{AARA}
\end{equation}
In the heavy quark limit, $m_h = m_h' = M_0 = M_0' = M = M'$, and
\begin{equation}
\left|{\partial {\bf V}' \over \partial {\bf v}'_h}\right|_{{\bf v}_l}
\to 1;\quad
\left| {\partial({\bf v}_h{\bf v}_l) 
\over\partial({\bf V}{\bf k})} \right|
\to {1\over m_l^3}
{\omega_1({\bf v}_l)
\over \omega_1({\bf u}_l)},
\label{AASA}
\end{equation}
with a similar result for the Jacobian involving primed variables. 
The remaining
dependence of the integral upon quark masses occurs via the momentum
${\bf k}$ as defined in Eq.~(\ref{AAF}):
\begin{eqnarray}
{\bf k} &&= {\bf p}_l - {\bf V}\left[ \omega_{m_l}({\bf p}_l)
- { ({\bf V}\cdot{\bf p}_l)\over 1 + V^0}\right]; \nonumber  \\
{\bf k}' &&= {\bf p}_l - {\bf V}'\left[ \omega_{m_l}({\bf p}_l)
- { ({\bf V}'\cdot{\bf p}_l)\over 1 + V'{}^0}\right].
\label{AAU}
\end{eqnarray}
The momenta ${\bf k}$, ${\bf k}'$ and ${\bf p}_l$ are related to each
other via Eq.~(\ref{AAU}).  They depend in turn on the external
velocities ${\bf V}$ and ${\bf V}'$, as well as $m_l$, {\it but they
  are independent of the heavy quark mass $m_h$.}  The remaining terms
in the Jacobians depend upon ratios of these three momenta to $m_l$,
and thus they also do not depend upon $m_h$.  The Wigner rotations of
the heavy quark become the identity in the heavy-quark limit.  Those
of the light degrees of freedom depend upon ${\bf k}/m_l$, which may
be non-negligible, as discussed below, but are independent of the
heavy quark mass.

We now investigate the contributions from the model to $\xi(w)$ as
$w\to 1$.  Since $\xi(w)$ is independent of the heavy quark mass, it
is sufficient to consider the case of elastic scattering from a heavy
pseudoscalar meson.  In the point form, all generators of the
homogeneous Lorentz group are kinematic, so that any component of the
current matrix element in any frame can be used to calculate invariant
quantities.  We therefore evaluate matrix elements of the charge
operator in the Breit frame, where ${\bf V} = -{\bf V}' =
{1\over2}{\bf A} = {1\over2}{\bf q}/M$, and ${\bf A}^2 = 2(w-1)$.

Before considering the general case of arbitrary light spin, we
present here a heuristic derivation of the minimum slope for light
spin ${1\over2}$.
With the normalization choice of Eq.~(\ref{AAA}),
the Isgur-Wise function becomes, in the heavy quark limit,
\begin{equation}
\xi(w) = (4\pi)^{-1} \int d{\bf k} \,
\phi_{0}^*(k')
D^{1\over2}_{{\bar\mu}_l'\mu_l}
[R_c(L_c({\textstyle{1\over2}}{\bf A}),-{\bf k}'/m_l)]
D^{{1\over2}{\dagger}}_{\mu_l{\bar\mu}_l}
[R_c(L_c(-{\textstyle{1\over2}}{\bf A}),-{\bf k}/m_l)]
\phi_{0}(k).
\label{AASh}
\end{equation}

For definiteness, we use an $S$-wave harmonic oscillator wave function
of the form $\phi_{0}(k) = \exp{(-k^2/2\beta^2)}/\sqrt{N}$.  The use
of an explicit wave function introduces model dependence, but for the
moment we are only interested in a heuristic derivation of the minimum
slope.

In the nonrelativistic limit, the slope of $\xi(w)$ as $w\to1$ is an
overall size effect due to the conventional form factor $F({\bf q}^2)
= 1 - {1\over6}{\bf q}^2R^2 + \cdots$.  In the harmonic oscillator
model, this is
\begin{equation}
  \label{ABKAA}
  F({\bf q}^2) = 1 - {1\over4}\left({m_l\over M}\right)^2
  {{\bf q}^2\over\beta^2} + \cdots,
\end{equation}
which yields [${\bf q}^2 = 2M^2(w-1)$]: 
\begin{equation}
  \label{ABLAA}
  \xi_{NR}(w) = 1 - {{1\over2}} {m_l^2\over\beta^2}(w-1) +
  \cdots. 
\end{equation}

While there are several contributions of relativistic origin to the
slope of $\xi(w)$, we consider only that of Wigner rotations for this
heuristic discussion.  In the limit $w\to 1$, the Wigner rotation angle
is, to lowest order in $k/m_l$,
\begin{equation}
  \label{ABLAB}
  \theta_R \to \left|{{\bf V}\times{\bf k}\over2 m_l}\right|,
\end{equation}
For the primed variables, the rotation axis is the same, and the
rotation angle has the same magnitude and opposite sign, with the
result that the product $D(\theta_R)D^{\dagger}(-\theta_R)$ can be
replaced by a single $D(2\theta_R)$.  The leading
contribution in powers of $(w-1)$ comes from the diagonal elements
$D^{1\over2}_{\pm{1\over2}\pm{1\over2}}(2\theta_R) = \cos\theta_R
\to 1 - {1\over2}\theta_R^2$.  For an $S$-wave harmonic oscillator
ground state, $\langle 0|({\bf V}\times{\bf k})^2|0\rangle = {\bf
  V}^2\beta^2$, and the net multiplicative contribution to $\xi(w)$ is
\begin{equation}
  \label{ABLAC}
  \xi_W(w) = 1 - {{1\over16}} {\beta^2\over m_l^2}(w-1) +
  \cdots. 
\end{equation}
Combining the factors in Eqs.~(\ref{ABLAA}) and (\ref{ABLAC}) yields
\begin{equation}
  \label{ABLAD}
  \xi(w) = \xi_W(w) \xi_{NR}(w) = 1 - {{1\over2}} {m_l^2\over\beta^2}(w-1) -
  {{1\over16}} {\beta^2\over m_l^2}(w-1) + 
  \cdots. 
\end{equation}
Note that the slope receives contributions both from factors
${m_l^2\over\beta^2}$ and ${\beta^2\over m_l^2}$.  In the extreme
nonrelativistic limit, $\beta\to 0$, and the slope becomes unbounded
due to the arbitrarily large geometric size proportional to
$1\over\beta$.  There exists a sum rule which provides an upper bound
on the slope~\cite{Voloshin,Bigi,Grozin,BoydLigeti}, implying that
this limit of the model is unphysical.  In addition, the form factor
contribution alone, which is present in a nonrelativistic calculation,
provides no lower bound to the slope.  The presence of the
relativistic correction via the Wigner rotation means that there is a
minimum slope of $\xi(w) = 1 - \rho^2(w-1)+\cdots$.  Minimizing the
coefficient of $(w-1)$ in Eq.~(\ref{ABLAD}) gives
\begin{equation}
  \label{ABLAE}
  \rho^2\ge {\sqrt{2}\over4},
\end{equation}
which differs by a factor $\sqrt{2}$ from the original Bjorken sum
rule~\cite{Bjorken,IWsum}.  However, the original sum rule receives
additional QCD corrections~\cite{Bigi,Grozin,BoydLigeti,BoydManohar}
which raise the lower bound of the slope to approximately 0.4.  Since
a direct-interaction quark model involves effective degrees of freedom
which incorporate QCD corrections implicitly, a comparison of
$\sqrt{2}/4=0.35$ to 0.4 can be taken as reasonable agreement, but
little can be said beyond this of a quantitative nature.  In any
event, there are other contributions beyond those of this heuristic
analysis.

We now turn to systems containing light degrees of freedom with
higher spin.  In this case there can be two or more distinct functions
$\xi(w)$.  For spin-one light spectators, this is well
known~\cite{IWbaryon,Georgi,Korner,Mannel}.  The number $n_\xi$ of
independent Isgur-Wise functions can be obtained by counting the
number of non-negative values of $\mu_l$, i.e., the number of unique
diagonal elements of the corresponding Wigner rotation.  For spin-zero
and spin-one mesons with $j_l={1\over2}$, and spin-${1\over2}$ baryons
with $j_l=0$, $n_\xi=1$.  For spin-one mesons with $j_l={3\over2}$,
and spin-${1\over2}$ baryons with $j_l=1$, $n_\xi=2$.  For now, we
consider only the matrix element for fully aligned spins, i.e.,
$j_{\rm hadron} = j_l+{1\over2}$.  The relevant Wigner rotation of
$2\theta_R$ is~\cite{BrinkSatchler}
\begin{equation}
  \label{ABLAF}
  D^{j_l}_{j_lj_l}(2\theta_R) = (\cos{\theta_R})^{2j_l}
  \to 1 - j_l\theta_R^2.
\end{equation}
Our heuristic minimization approach yields the following behavior for
$\xi(w)$: 
\begin{equation}
  \label{ABLAG}
  \xi_W(w) = 1 - {{\sqrt{j_l}\over2}} {\beta^2\over m_l^2}(w-1) +
  \cdots. 
\end{equation}

For spins which are not fully aligned, the expression for the minimum
slope becomes more complicated but is always less than $\sqrt{j_l}/2$.

We now consider arbitrary light spin with full relativistic effects
For spin-${1\over2}$ spectators, Le~Yaouanc {\it et al.}\ find that a
careful analysis yields the following result for $\rho^2_{\rm min}$:
\begin{eqnarray}
  \label{ABLAH}
  \rho^2_{\rm min} &= & \rho^2_{\rm space} + \rho^2_{\rm Wigner} +
  \rho^2_{\rm quark}
  \nonumber \\
  &=& {1\over3} + {1\over6} + {1\over4},
\end{eqnarray}
where $\rho^2_{\rm space}$ contains contributions due to boosts of
wave function argument and Jacobians for variable transformations,
$\rho^2_{\rm Wigner}$ is the contribution from spectator Wigner
rotations, and $\rho^2_{\rm quark}$ is a normalization effect from the
heavy quark current.  They find that this minimum corresponds to
arbitrarily large internal momenta ($\beta\to\infty$ in the harmonic
oscillator picture).  Given this result, one expects no change in the
value of $\beta$ which minimizes $\rho^2$ when the contribution of the
Wigner rotation to $\rho^2$ is altered by a change in its coefficient
from ${1\over2}$ to some other value of $j_l$.  We thus expect that
for general $j_l$ and fully aligned spins,
\begin{equation}
  \label{ABLAK}
  \rho^2_{\rm min} = {1\over3} + {j_l\over3} + {1\over4}.
\end{equation}

The slope grows with $j_l$ for large $j_l$, in agreement with Falk.
However, it is difficult to make a detailed comparison between the
tensor/sum-rule approach and quark models of the sort described here.
In the tensor/sum-rule approach the minimum slope vanishes when
$j_l=0$; this would correspond to $\rho^2_{\rm space} = \rho^2_{\rm
  quark} = 0$ in the quark model.  In the nonrelativistic sense that
$\rho^2_{\rm space}$ represents a geometric radius, this limit
corresponds to $\beta\to\infty$.  But we have already seen that there
are contributions to $\rho^2_{\rm space}$ which are either independent
of $\beta$, or which have a non-vanishing limit as $\beta\to\infty$.
As noted by Le~Yaouanc {\it et al.}, the additional non-vanishing
contributions $\rho^2_{\rm space}$ and $\rho^2_{\rm quark}$ correspond
to the {\it compositeness} of the hadron as described by the quark
model.  By contrast, the analyses based upon a tensor representation
describe the hadron as a fully factorized product of heavy quark and
spectator degrees of freedom.  In a sense they correspond to the wave
function at zero separation in the quark model, i.e., the limit of
arbitrarily large internal momenta for which $\rho^2_{\rm Wigner} =
{1\over6}$.  This reasoning may represent a partial connection between
the quark model and the tensor/sum-rule approaches, but the difference
between ${1\over6}$ in the former and ${1\over4}$ in the latter
remains unexplained.

In summary, the effect of relativity attributed to the minimum slope
of the Isgur-Wise function as obtained from sum rules can be seen from
the viewpoint of a simple model to arise naturally from Wigner
rotations of the light degrees of freedom.  Direct-interaction models
have other contributions to the minimum slope which are of
relativistic origin, but they are independent of the light spin and
thus have no counterpart in the HQET sum-rule treatment.  It may well
be that any composite model of a heavy hadron will yield a minimum
slope of $\xi(w)$ which exceeds the sum-rule value.

The author wishes to thank N.~Isgur for asking about the source of a
minimum slope in models of the sort considered here, G. C. Boyd and W.
Roberts for helpful comments, and O. P\`ene for correspondence
concerning Refs.~\cite{LeYaouanc1,LeYaouanc2,LeYaouanc3}.  This work
was supported in part by the U.S. National Science Foundation under
Grant PHY-9319641.

\mediumtext


\begin{references}
\bibitem{IW} N. Isgur and M. B. Wise, in {\it B Decays}, edited by
  S. Stone (World Scientific, Singapore, 1991), p.~158.
\bibitem{Bjorken} J. D. Bjorken, invited talk at Les Reoncontre de
  Physique de la Valle d'Aoste, La Thuile, Italy [SLAC Report No.\
  SLAC-PUB-5278, 1990 (unpublished)]; J. D. Bjorken, I. Dunietz and
  J. Taron, Nucl.\ Phys.\ {\bf B371}, 111 (1992).
\bibitem{IWsum} N. Isgur and M. B. Wise, Phys.\ Rev.\ D {\bf 43}, 819
  (1991).
\bibitem{Falk} A. F. Falk, Nucl.\ Phys.\ {\bf B378}, 79 (1992).
\bibitem{Close} F. E. Close and A. Wambach, Nucl.\ Phys.\ {\bf B412},
  169 (1994).
\bibitem{LeYaouanc1} A. Le Yaouanc, L. Oliver, O. P\`ene and J.-C. Raynal,
  Phys.\ Lett.\ {\bf B365}, 319 (1995).
\bibitem{LeYaouanc2} A. Le Yaouanc, L. Oliver, O. P\`ene and J.-C. Raynal,
  Phys.\ Lett.\ {\bf B386}, 304 (1996).
\bibitem{LeYaouanc3} V. Mor\'enas, A. Le Yaouanc, L. Oliver, O. P\`ene
  and J.-C. Raynal, Phys.\ Lett.\ {\bf B365}, 315 (1996).
\bibitem{BakamjianThomas} B. Bakamjian and L. H. Thomas, Phys.\ Rev.\
  {\bf 92}, 1300 (1953).
\bibitem{KeisterPolyzou} B. D. Keister and W. N. Polyzou, Adv.\ Nucl.\
  Phys.\ {\bf 20}, 225 (1991).
\bibitem{BKheavy} B. D. Keister, Phys.\ Rev.\ D {\bf 46}, 3188 (1992).
\bibitem{ISGW2} D. Scora and N. Isgur, Phys.\ Rev.\ D {\bf 52}, 2783
  (1995). 
\bibitem{Voloshin} M. B. Voloshin, Phys.\ Rev.\ D {\bf 46}, 3062
  (1992).
\bibitem{Bigi} I. Bigi, M. Shifman, N. G. Uraltsev and A. Vainshtein,
  Phys.\ Rev.\ D {\bf 52}, 196 (1995).
\bibitem{Grozin} A. G. Grozin and G. P. Korchemsky, Phys.\ Rev.\ D
  {\bf 53}, 1378 (1996).
\bibitem{BoydLigeti} C. G. Boyd, Z. Ligeti, I. Z. Rothstein and
  M. B. Wise, Phys.\ Rev.\ D {\bf 55}, 3027 (1997).
\bibitem{BoydManohar} C. G. Boyd, B. Grinstein and A. V. Manohar,
  Phys.\ Rev.\ D {\bf 54}, 2081 (1996).
\bibitem{IWbaryon} N. Isgur and M. B. Wise, Nucl.\ Phys.\ B {\bf 348},
  276 (1991).
\bibitem{Georgi} H. Georgi, Nucl.\ Phys.\ B {\bf 348}, 293 (1991).
\bibitem{Korner} J. G. K\"orner, Nucl.\ Phys.\ {\bf B} (Proc.\ Suppl.)
  {\bf 21}, 366 (1991)
\bibitem{Mannel} T. Mannel, W. Roberts and Z. Ryzak, Nucl.\ Phys.\ B
  {\bf 355}, 38 (1991).
\bibitem{BrinkSatchler} D. M. Brink and G. R. Satchler, {\it Angular
    Momentum, 2nd Edition} (Oxford University Press, London, 1968).
\bibitem{Close} F. E. Close, {\it An Introduction to Quarks and
  Partons} (Academic Press, New York, 1979), and references therein.
\end{references}
\end{document}